\documentclass[a4paper,11pt]{article}
\usepackage{jinstpub} 

\title{A compact and inexpensive device for biasing beam steerer electrodes}

\author{P. Schury}
\affiliation{Wako Nuclear Science Center (WNSC), Institute of Particle and Nuclear Studies (IPNS), High Energy Accelerator Research Organization (KEK), Wako, Japan}

\emailAdd{peter.schury@kek.jp}

\abstract{We introduce a compact design for an electronic device capable of superposing 8 bias voltages up to $\pm$96~V on a voltage of up to 5~kV.  Such a device has been implemented to provide biasing of a beam steerer-shifter system for use in aligning an ion beam with the optical axis of a multi-reflection time-of-flight mass spectrograph.  The device features a 16-bit digital interface both to set and to monitor the applied voltages, and can be built for less than \$500~USD.  By selection of higher isolation components, the design concept can be extended to allow operation up to 20~kV.
}


\arxivnumber{1234.56789} 

\begin{document}
\maketitle
\flushbottom

\section{Introduction}
\label{sec:intro}

\par The ability to steer or axially shift an ion beam is crucial in most ion optical systems.  If a steering element were to axially accelerate or decelerate the ion beam, it would act as a focusing element and induce an axial displacement dependent steering.  To prevent such undesirable behavior, the steering element must be at the same nominal potential as the preceding optical element. Often, this requires such a steering system to have a nominal potential exceeding 1000~V.  

\par A common approach is to bias each element with an independent high-voltage supply. However, this quickly becomes prohibitively expensive.  A common alternate solution, often employed when the beam line itself is on some high-voltage potential, is to use a high-voltage isolation transformer to ``float" an array of digital-to-analog (DAC) circuits on the beam potential.  In cases where the beam line is biased at some high-voltage potential, such a DAC system is generally a minor addition to other devices (vacuum pumps and gauges, ion detectors, etc.) which would be powered from such an isolation transformer.  However, in a case where the beam line is not biased, it is an expensive and bulky option that must also be accompanied by optical isolation equipment for digital communication with the DAC.

\par
\begin{figure}[t]
 \includegraphics[width=0.98\textwidth]{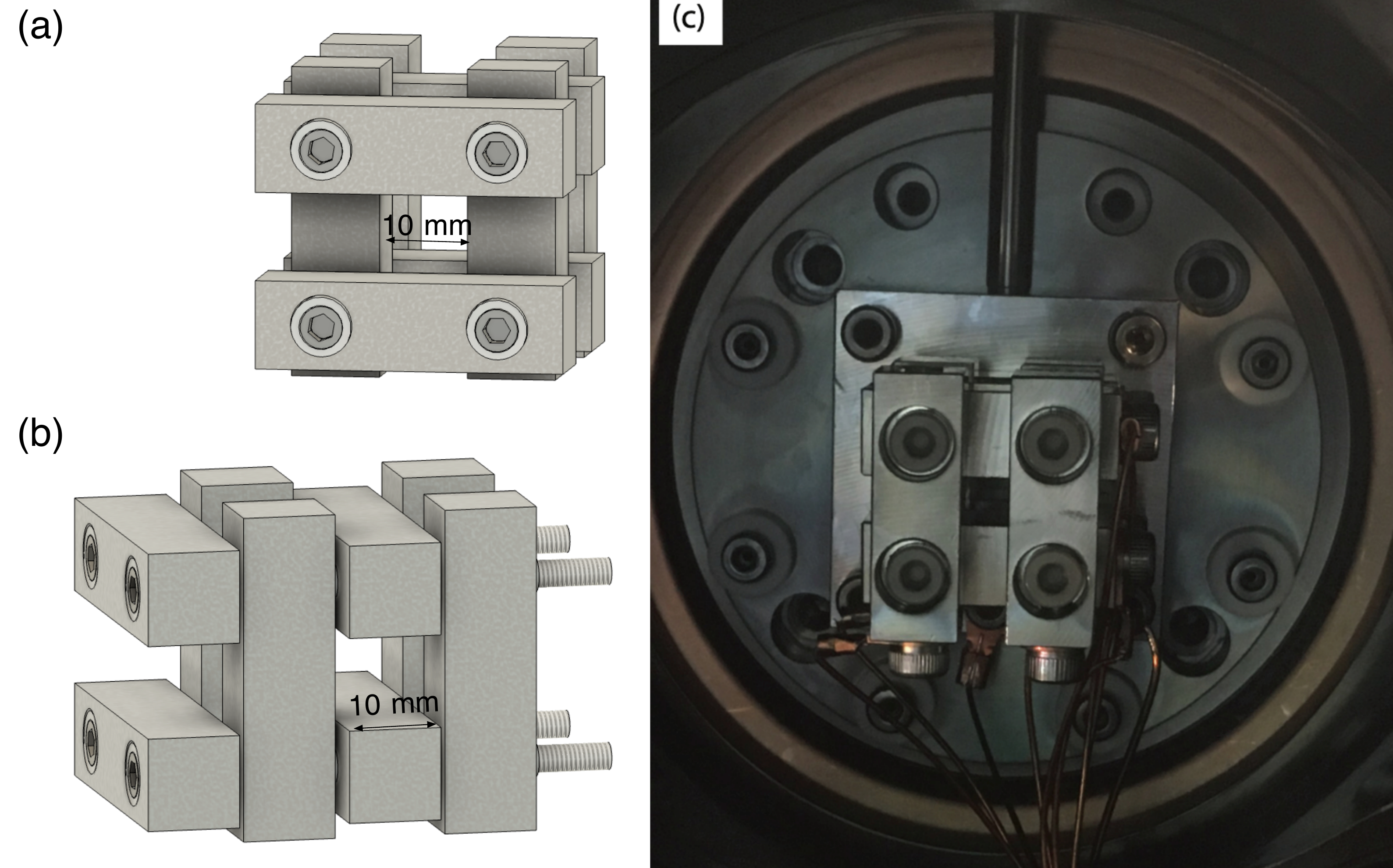}
\caption{(a) and (b) Rough schematic drawings showing the ion steerer-shifter layout. (c) Photograph of ion steerer-shifter as implemented in our systems.  To steer/shift the ions without introducing focusing requires that the mean potential of the steerer should match that of the preceding optical element.  As there are eight electrodes, independent biasing would require eight high-voltage modules.}
\label{figElectrodes}
\end{figure}

 In the implementation of our multi-reflection time-of-flight mass spectrographs (MRTOF-MS) \cite{Schury2014}\cite{MarcoZD}\cite{KISS-MRTOF} at the RIKEN Nishina Center for Accelerator Based Science (RNC), we had a need for the biasing of eight-element beam steerer-shifters (see Fig.~\ref{figElectrodes}).  This made the use of multiple, independent high-voltage supplies uneconomical.  Because the beam line itself is not biased in our system, and space is highly limited, a compact solution was particularly desirable.  To meet these needs, we have designed an inexpensive, compact, high-precision steerer-shifter bias system.

\par Using modern digital isolators and isolated DC-DC power supplies, we have built and tested a device with eight  $\pm$96~V output channels capable of floating on as much as 5~kV.  A relatively inexpensive Raspberry Pi \cite{Raspi} single board computer (SBC) is used to provide an SPI interface for controlling the device and an Ethernet connection for a web-based user interface, enabling remote control and real-time monitoring of the steerer voltages without the need for dedicated client software.  In principle the Rspberry Pi could be replaced by any number of other SBCs or even a micro-controller such as from the STM32, ESP32, or Raspberry Pi Pico lines of devices.  Each channel employs a 16-bit DAC to set the voltage and a 16-bit ADC for monitoring.  The design, as built, cost less than \$500 USD.  Using an alternative choice of digital isolator and DC-DC power supply, the device topology should be able to be extended to allow floating on up to 20~kV.

\section{Design}
\label{secDesign}

\par To implement such a device as required for steering and shifting our ion beam requires solutions for three distinct problems: provide power to devices floating on $V_\textrm{iso}>$1~kV, provide digital communication with these devices floating on the voltage $V_\textrm{iso}$, and implement high-voltage op-amps to provide sufficient steering voltages.  Modern DC-DC converters are readily available in compact form-factors with isolation ratings exceeding 5~kV$_\textrm{DC}$; Recom's RHV line even offers 20~kV$_\textrm{DC}$ isolation rating \cite{RecomRHV}.  Similarly, many modern digital isolators use capacitive digital isolation in which the input signal is typically encoded via on–off keying and transmitted across an integrated silicon dioxide dielectric barrier using on-chip metal–oxide–metal capacitors, followed by demodulation on the secondary side \cite{Zhao2024}, which have isolation ratings exceeding 5~kV$_\textrm{DC}$ and can operate with digital rates far exceeding 10~Mbps.  For the high-voltage op-amp, recently a variety of compact and inexpensive op-amps supporting rail voltages up to or exceeding 100~V have become available, such as ADA4700-1 (100~V rated), LTC6090 (140~V rated), and ADHV4702-1 (220~V rated).

\par
\begin{figure}[t]
 \includegraphics[width=0.58\textwidth]{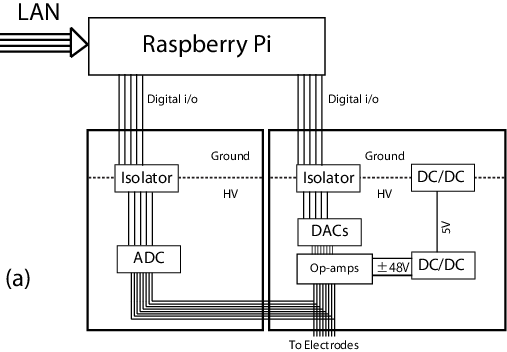}
 \includegraphics[width=0.41\textwidth]{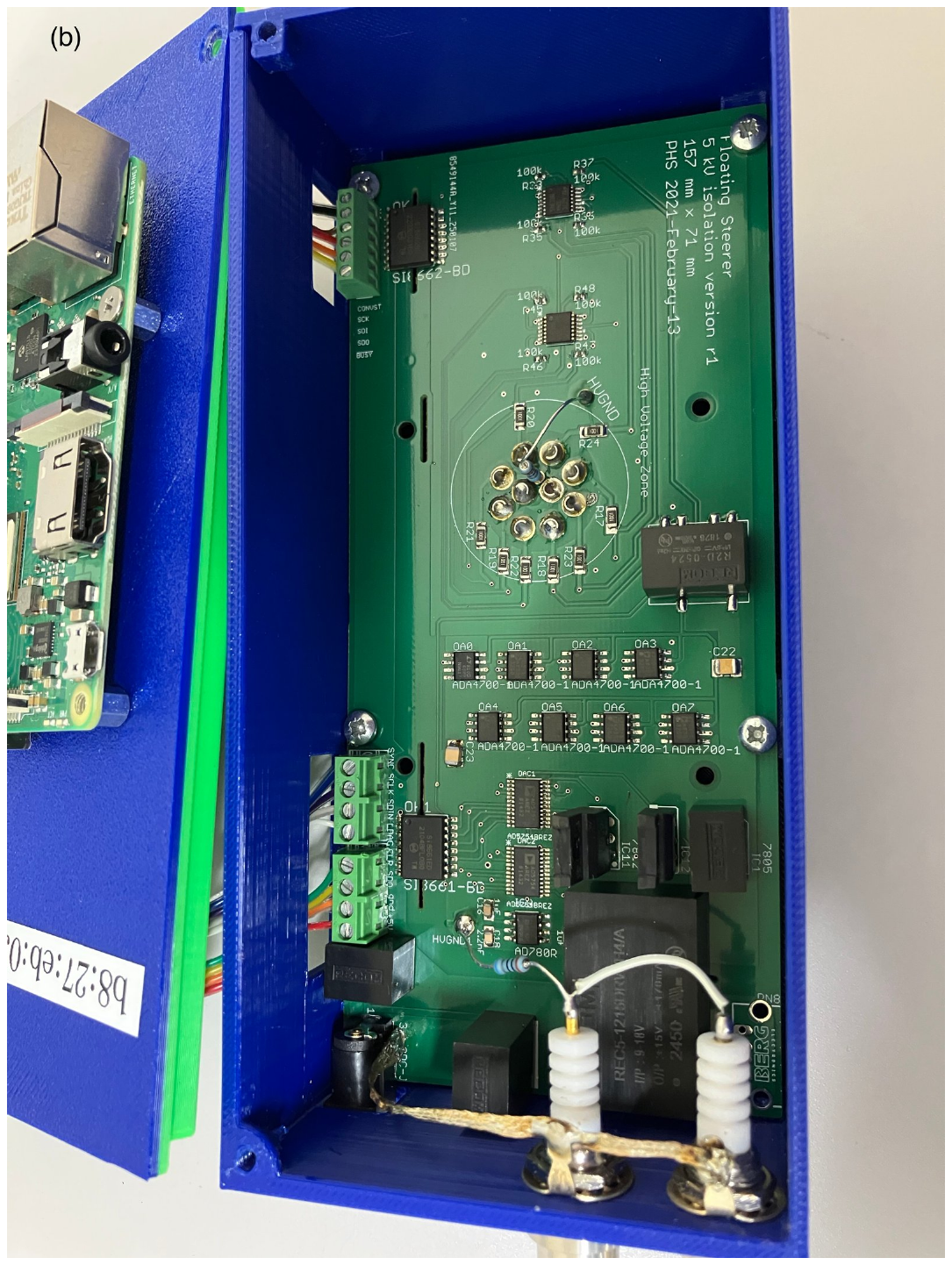}
\caption{(a) Schematic overview of the floating ion steerer electronics design. (b) Photograph of the floating ion steerer electronics as implemented.  The circuitry is mounted inside a 3D printed PETG box for safety; the Raspberry Pi is mounted on the lid, with digital and 5~V power connected by ribbon cables via the small windows seen at the left in the photo.  Two SHV connectors allow for applying the floating potential in a daisy-chain configuration.  The hidden underside of the PCB contains bypass capacitors, one 5V-to-48V DC-DC converter, and an LTC1858 analog-to-digital converter chip. In versions with higher steerer voltage capability, more 5V-to-48V DC-DC converters are employed in a stack.}
\label{figDevicePhoto}
\end{figure}

 In this section we will discuss our implementation of modern DC-DC converters, digital isolators, and high-voltage op-amps, along with the other components required to implement the device.  The general topology is shown schematically in Fig.~\ref{figDevicePhoto}(a), additional design details may be made available by the authors subject to institutional and export control regulations.  A fully-implemented version of the device is shown in Fig.~\ref{figDevicePhoto}(b), including the 3D-printed enclosure to provide safety. The entire device is 165~mm x 75~mm and 55~mm in height.  The use of an MS-10 multi-pin connector allows the device the be fastened directly to the beam line at feedthrough, further increasing the compactness.

\subsection{Power distribution and isolation}

\par As previously mentioned, the traditional means to power devices on an elevated electric potential is to use an isolation transformer.  Isolation transformers rated for 10~kV isolation and capable of providing >1~kW of power are readily available from multiple manufacturers. They provide A/C mains power, making them easy to use with off-the-shelf equipment, but they tend to be physically large and expensive.  Very small and inexpensive DC-DC converters, meanwhile, are offered by many manufacturers with output isolation rated up to 20~kV, although with typical maximum output power of $\le$10~W; they also provide a DC voltage rather than an A/C signal.  It is common practice to employ DC-DC converters to provide a galvanic break in order to prevent ground loops and to protect delicate circuitry, in addition to their use in efficiently stepping-up and stepping-down DC voltages.  We have taken advantage of these compact and inexpensive devices by using commercial DC-DC converters for provision of isolated DC power to equipment operating on elevated potentials.

\par The concept of using isolated DC-DC converters to provide low-voltage power to an isolated circuit on high-voltage is not novel \cite{Hitzemann2022}, however the literature discussing it is somewhat limited.  The datasheets for commercial devices define the isolation voltage limit in terms of hi-pot testing for between 1~s and 2~minutes, while stating the ``working voltage'' as being as much as an order of magnitude lower.  However, after testing several DC-DC converters from a variety of manufacturers we have yet to see one suffer breakdown near the listed isolation limit even after years of constant operation.  As such, we are now sufficiently confident in the reliability of the method to publish our technique.

\par To provide power to the ion steerer's high-side electronics, we made use of a TEN 8-1223 DC-DC converter from Traco Power, which features 1.5~kV isolation and is capable of 8~W output power at $\pm$15~V in our first design; later we changed to REC6-1215D/R10 from Recom which offers a 6~W output power at $\pm$15~V with an isolation rating of 10~kV to allow higher voltage to be applied to our steering unit.  For the various circuitry on the high-voltage side of the device, the DC-DC converter drives multiple voltage regulators, a precision voltage reference (AD780R), and a set of R2D-0524 step-up DC-DC converters.  The $\pm$15~V-output DC-DC converter was chosen to provide sufficient head-space to use inexpensive 7812 and 7912 series linear regulators to produce $\pm$12~V in order to power the DAC with the clean rail voltages while still allowing a $\pm$10~V output range.

\begin{figure}[tbh]
 \includegraphics[width=0.9\textwidth]{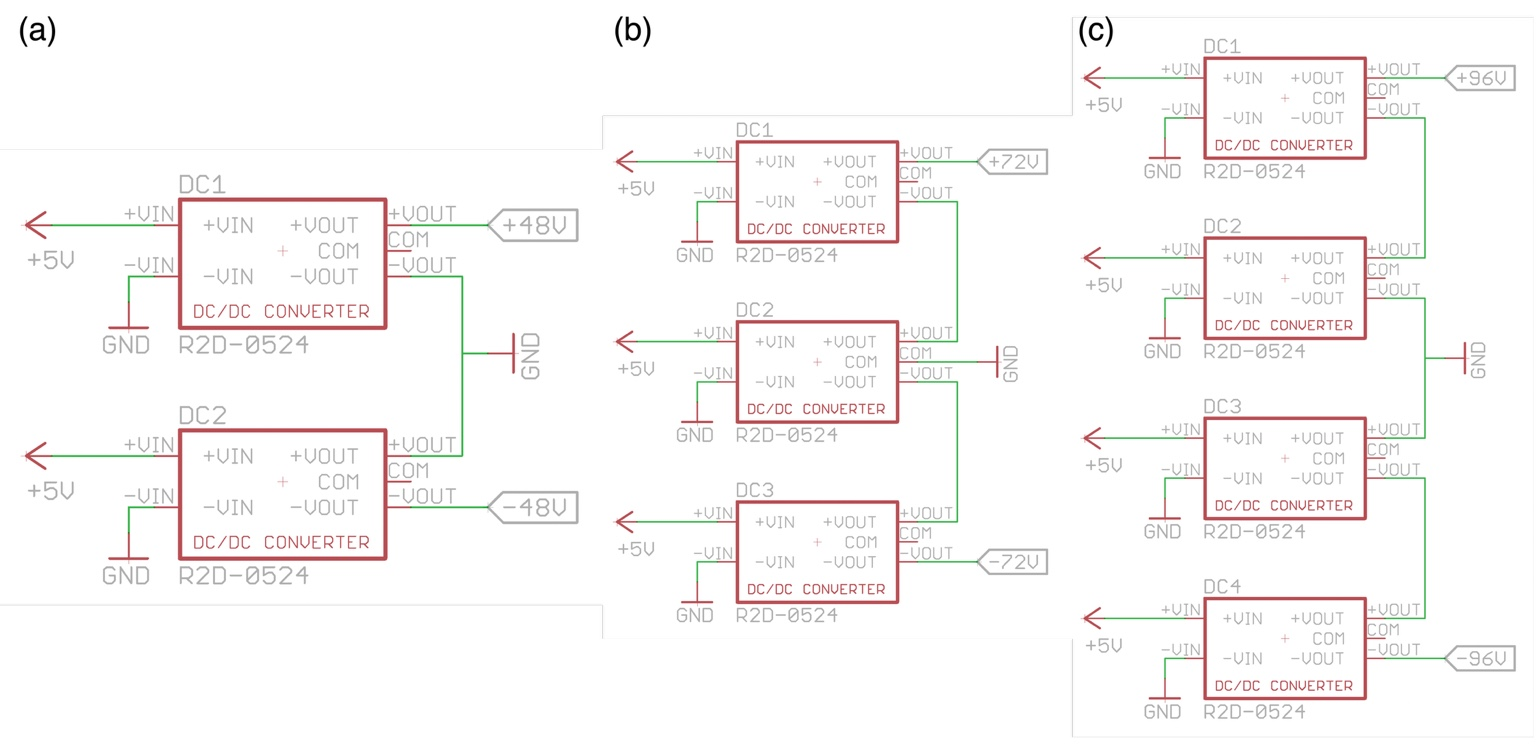}
\caption{Schematic demonstration of how to use stacks of R2D-0524 to generate (a) $\pm$48~V, (b) $\pm$72~V, and (c) $\pm$96~V rails for high-voltage op-amps.  In this schematic GND refers to the high-voltage floating ground.}
\label{figDCDC}
\end{figure}

\par To provide the high-voltage rails for the op-amps as they are inexpensive, compact, and robust we selected the DC-DC converter R2D-xx24.  We opted to use the R2D-0524 with 5~V input powered via the highly-efficient R-785.0-0.5 switching mode version of the 7805 voltage regulator as it would result in less power draw from the primary DCDC than using R2D-1224 powered by the 7812 linear regulator.  The R2D-0524 provide $\pm$24~V output voltages that are isolated from the input voltage, so by tying the negative output to the floating ($i.e.$ high-voltage) ground the positive output will be +48~V.  Similarly, if the positive output is tied to floating ground, the negative output will be -48~V.  In this way, we can generate the $\pm$48~V for the high-voltage op-amps.  For the $\pm$72~V version, three of these DC-DC converters can be used, while four are required for the $\pm$96~V version; see Fig.~\ref{figDCDC}.

\subsection{Digital signal isolation}

\par In addition to isolated power, digital communications signals must be transferred to the floating electronics.  This is accomplished through digital isolators.  Until recently, the only option for such isolated digital signal transmission was via optical couplings.  While fibre optic devices capable of extreme high-speed communication exist, small-scale optocouplers are typically capable of 10~Mbps operation at most; such small-scale optocouplers also have fairly limited maximum isolation voltages.  Additionally, they require external components and draw considerable supply current.  Starting in the early 2010's, however, digital isolators based on giant magnetoresonance (GMR) came to market.  These devices are capable of isolating digital signals with rates of 100~Mbps or faster, owing to signal skews on the level of a few nanoseconds, while not requiring external components and drawing very limited supply current.  Our initial implementation made use of such GMR-based digital isolators (IL260 and IL262, rated for 2.5~kV$_\textrm{rms}$ isolation and data rates up to 110 Mbps) for communications with the DAC and ADC devices floating on the high voltage.  

\par More recently, however, capacitive digital isolation with ever higher isolation barrier strength has become ever more commonplace.  Such devices are capable of operating at data rates of 100~Mbps or more.  The dielectric barriers used in these devices is generally silicon dioxide of at least 10~$\mu$m thickness, giving them a breakdown withstand voltage 10~kV or more; the effective isolation voltage is largely limited by creepage distance of the package.  For our more current designs, we have opted to use SI8661 and SI8662 wide-SOIC isolators with isolation voltage rating listed as 5~kV$_\textrm{rms}$.  To maximize the creepage distance and thereby the effective isolation voltage, as seen in Fig.~\ref{figDevicePhoto}(b), we have cut a slot under and beyond the frame of the isolators.

\par As with the DC–DC converters, detailed information on the onset of dielectric failure under long-term operation is generally not provided in manufacturer documentation. For the SI866x wide-SOIC devices used here, the isolation rating is specified as ``5~kV$_\textrm{rms}$ for 1~minute’’, while the recommended working voltage is given as 1200~V, corresponding to an expected lifetime on the order of 60~years. These working voltage ratings are typically derived under conservative assumptions appropriate for safety-critical applications (e.g.\ medical or industrial systems), where extremely low failure probabilities over multi-decade operation are required.  In the present experimental context, only a limited number of devices are deployed, and operational lifetimes are typically much shorter. Under these conditions, the acceptable balance between performance and long-term reliability can differ from that assumed in such conservative specifications. Empirically, we note that operation of these digital isolators at voltages exceeding the specified working voltage over periods of several years has not resulted in observed failures within our limited sample size and duration of operation. While this does not constitute a systematic lifetime study, it provides practical evidence that the devices can tolerate such conditions in the specific operating regime considered here.

\subsection{Digital interface and high-voltage amplification}

\par To set the steering voltages, a pair of daisy-chained AD5754 (Analog Devices) digital-to-analog-converters (DAC) are used.  These chips offer 4 bipolar output with 16-bit precision.  Programming is performed via a 4-wire SPI interface with a maximum data rate of 30 Mbps.  The aforementioned AD780R precision reference provides a 2.500~V reference which allows the DAC outputs to be programmed in a range of $-10~V$ to $+10~V$.  

\par The bipolar outputs from the DAC allows for a maximally simple implementation of the amplification stage, as it obviates the need to shift the midpoint and ensures that the device powers on with the steering functionally disabled.  We have tested multiple amplification stage designs, progressively using three high-voltage op-amps with higher and higher available ranges: ADA4700-1, LTC6090, and ADHV4702-1.  In each design we implemented an inverting topology -- with the $\pm$48~V-range ADA4700-1 we used $g=-5$, with the $\pm$72~V-range LTC6090 we used $g=-7.5$, and with the $\pm$96~V-range ADHV4702-1 we used $g=-10$.

\par To ensure that the DAC and amplifiers are working properly, an analog-to-digital converter (ADC) system was implemented for monitoring.  An eight-channel, 16-bit SAR ADC (LTC1856 from Linear Technology Corporation) has been used for this purpose.  This ADC can be powered from a unipolar 5~V supply, while still being able to digitize signals within a $\pm$10~V bipolar input range and also includes its own onboard precision voltage reference.  

\par To safely digitize the voltages applied to the steerer, which may span as much as $\pm$96~V, we use a resistive divider to reduce the maximum range to $\leq\pm$10~V.  To utilize the full digitization range while also balancing the current drawn by the divide against noise generated by high-resistance paths, we opt for a high-side resistance of 100~k$\Omega$ and a low-side resistance of 20~k$\Omega$, 15~k$\Omega$, or 10~k$\Omega$ for the $\pm$48~V, $\pm$72~V, $\pm$96~V versions, respectively.  Because the LTC1856 has an input impedance of 31~k$\Omega$, the connection to the ADC is buffered by an op-amp  (MC33079, On Semiconductor) in voltage-follower configuration to produce a low-impedance connection to the ADC input. 

\par Based on the 16-bit resolution of the ADC and DAC used in the device, one least significant bit (LSB) corresponds to approximately 5~mV for both the set and monitor values. In practice (see Fig.~\ref{fig2DScan}), voltage control at the 1~V level is generally sufficient for beam alignment. The monitor readback shows typical agreement with the programmed value to better than 10~LSB and long-term stability better than 5~LSB$_\textrm{rms}$ over several weeks of observation. Use of lower temperature-coefficient resistors could further improve long-term stability, although this was not required for the intended application presented here.

\section{Performance results}

\par While the DC isolation ratings are giving for only a 1~second or 60~second test pulse in most cases, we have found that these devices generally are fully capable to operate at or near their rated limits for long terms.  A first version of the device used TEN 8-1223 DC-DC converters and IL262 digital isolators, both rated for 1.5~kV$_\textrm{DC}$ isolation, and they worked without fail at $V_\textrm{iso}\approx$1.2~kV for more than 2~years of service life before being replaced to allow higher voltages.  With increasing voltage requirements, we have now implemented the device presented herein which uses REC6-1215D DC-DC converters and SI866x digital isolates, both with isolation ratings of 5~kV$_\textrm{DC}$; they have been operated with our full-scale MRTOF-MS devices\cite{Schury2014}\cite{MarcoZD} at $V_\textrm{iso}\approx$3~kV for more than 3 years of service life without failure.

\begin{figure}[bt]
 \includegraphics[width=0.98\textwidth]{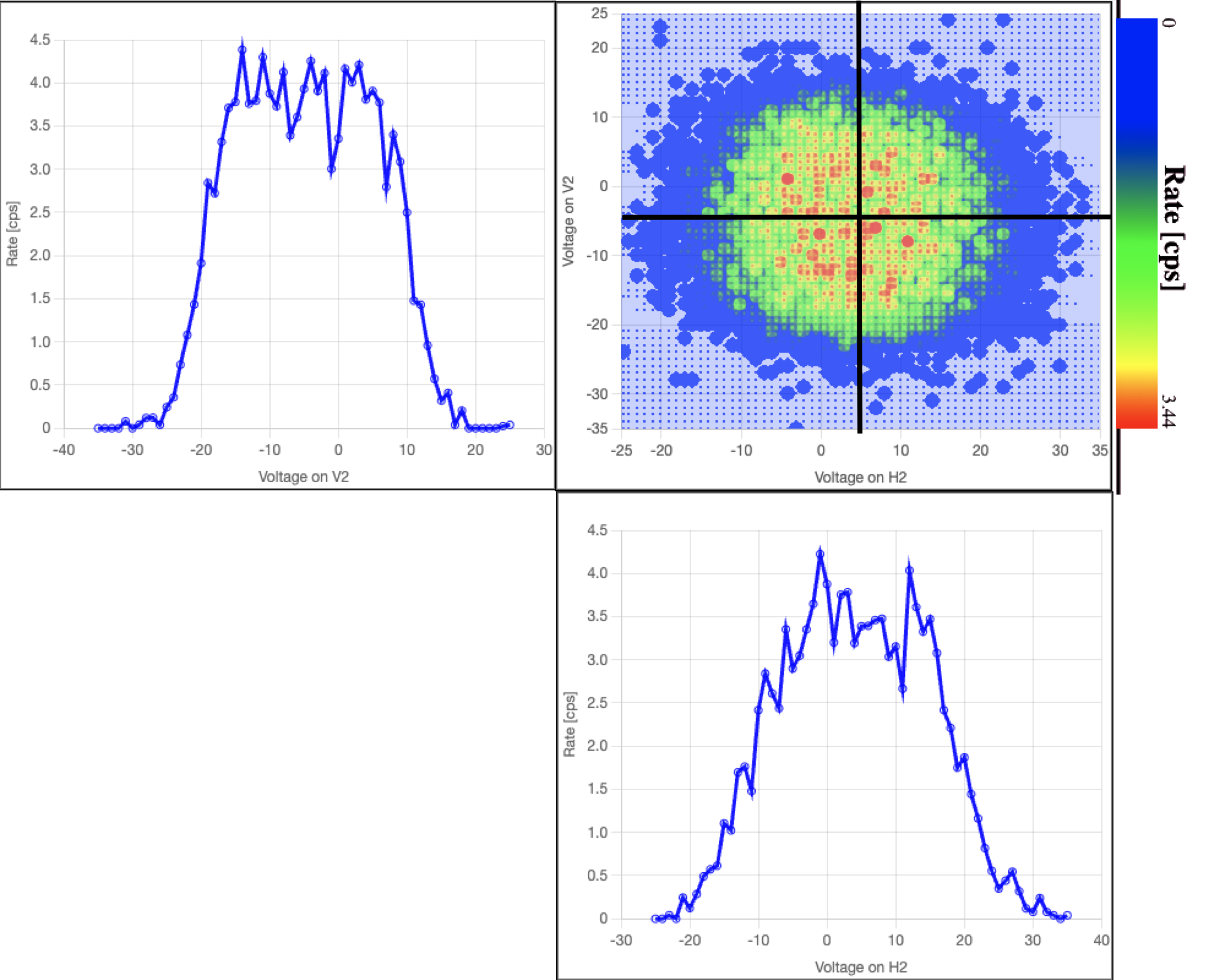}
\caption{Result of a two-dimensional steering scan.  The scan was performed as a so-called ``snake-scan'' where for V2$_i$ if the values of H2 were incremented, then for V2$_{i+1}$ the values of H2 were decremented.  This ensures that no false counts from voltage transits can appear in the first value of a scan line.  The one-dimensional scans of H2 at optimal V2 and V2 at optimal H2 are also given.  See text for naming conventions.  Common-mode voltage was $-1.25$~kV.}
\label{fig2DScan}
\end{figure}

\par While we have not experienced breakdown of the DCDC converters or digital isolators, we have had conditions where there can be discharges at the electrodes.  This was more common in the early period of operation, where some small sharp points in the structure can discharge more easily to a nearby wire, the grounded vacuum chamber, or an adjacent electrode.  After some time, these sharp points are removed naturally through micro-discharges.  However, we did find that occasionally larger discharges at the electrodes would result in a significant, if short-lived, high-voltage current being drawn through the steerer electronics and causing damage particularly to the relatively sensitive DAC chips.  To ameliorate this problem, we added a 1~M$\Omega$ resistor between the HV input and the PCB high-voltage bias connection and a 1/8-W 100~$\Omega$ resistor between the HV plane of the PCB and the HV pass-through to the steerer support frame.  The former limits the current which can flow through the circuitry during a discharge event, while the latter acts as a fuse -- if more than 50~mA is drawn through it, it will quickly burn out and create an open circuit, protecting the delicate components.  Since installing these safety resistors, no further units have been damaged by discharges in the electrodes. Such discharges in the electrodes have occurred on occasion, most commonly as a result of vacuum excursions (temporary degradations in vacuum level), and have required the 100~$\Omega$ resistor to be replaced.

\par Finally, to demonstrate the use of the steerer, we present the results of a steerer optimization measurement.  For this, we used $^{85}$Rb$^+$ ions from a thermal ion source at our half-scale MRTOF-MS\cite{KISS-MRTOF}.  The ions were accumulated and stored in our ion trap front end \cite{SchuryWB2} before being sent to the MRTOF-MS via the beam steerer-shifter shown in Fig.~\ref{figElectrodes}.  To increase the steering sensitivity, the ion time-of-flight was set to 1600~laps ($t\approx$25~ms corresponding to an effective flight path of $l\approx$800~m), making precise alignment with the optical axis critical.  With the first vertical and horizontal steerer electrodes biased to V1=2~V and H1=$-2$~V (relative to their common-mode high-voltage potential of $-1.25$~kV) we performed a two-dimensional scan of the voltages applied to second pair of steerer electrodes (V2 and H2) with $\Delta V$=1~V steps.  The result is shown in Fig.~\ref{fig2DScan}.  In this case, the MRTOF-MS reflection chamber had been well-aligned with the ion trap axis and the steering voltages were rather small; this is not always the case.

\section{Summary and outlook}
\par We have described the use of modern DC-DC converters and digital isolators (both GMR-based and capacitive to power and communicate with electronics which are floating on a high-voltage potential up to 10~kV.  We have implemented this to construct an 8-channel $\pm$48~V range power supply that can float on potentials up to 1.5~kV for use in biasing ion beam steering elements.  The device can be built for less than \$500~USD, making it a highly economical option.

\par We have observed that by taking care to maximize creepage distances it is possible to safely operate isolated DC-DC converters and digital isolators well-above their specified ``working voltage'' and near their listed maximum isolation voltage for durations of years.  

\par With appropriate modifications, the steerer electronics may be adapted for operation at significantly higher floating potentials. For example, wireless power transfer schemes ($e.g.$\ inductive coupling) and wireless communication could, in principle, eliminate the need for galvanic connections across the isolation barrier. While such approaches introduce additional design challenges, they may offer a pathway toward operation at substantially higher voltages than those demonstrated here. Exploration of such implementations is left for future work.

\acknowledgments

\par The authors wish to acknowledge the support of the Nishina Center for Accelerator Sciences.  This work was supported by the Japan Society for the Promotion of Science KAKENHI (grants \#2200823, \#24224008 and \#24740142).




\end{document}